\documentstyle[12pt]{article}
\begin{document}
\title{Physics from Bose-Einstein correlations in high energy multiparticle
production}
\author{Kacper Zalewski\thanks{Supported in part by the KBN grant
2P03B 08614}\\
Institute of Physics of the Jagellonian University\\ and\\ Institute of
Nuclear Physics, Krak\'ow, Poland}
\maketitle

\abstract{Bose-Einstein correlations are being exploited to obtain information
about the structure of the sources of hadrons in multiple particle production
processes. In this paper the principles of this approach are described and some
of the controversies about their implementation are discussed.}
\vspace{2cm}

\section{Introduction}

Bose-Einstein correlations among the momenta of identical particles produced in
a high energy multiple particle production process yield information about
the structure of the source of hadrons in the process. This has been pointed
out in the very first paper on these correlations \cite{GGL}. In a recent
Physics Report issue \cite{WIH} Wiedemann and Heinz write: {\it Two-particle
correlations provide the only known way to obtain directly information about
the space-time structure of the source from the measured particle momenta.}
Thus the problem of extracting as well as possible the information about the
source from the measured correlations is of great importance. It is, however,
not an easy problem. One of the founders of this field of research, G.
Goldhaber wrote in 1990 \cite{GOL}: {\it What is clear is that we have been
working on this effect for thirty years. What is not as clear is that we have
come much closer to a precise understanding of the effect.} It is unlikely that
the progress made during the last ten years would make G. Goldhaber give a much
more optimistic view.

There is a great variety of multiple particle production processes. In order to
illustrate this point we will describe two well-known cases. In an
electron-positon annihilation at high energy, at first usually a
quark-antiquark pair is produced. These partons radiate gluons. The gluons
radiate further gluons, or go over into quark-antiquark pairs. After some steps
of this cascade, in a process known as hadronization and not well understood,
the partons combine into colour-neutral hadrons. The hadrons are collimated
into two narrow jets pointing in opposite directions along the same straight
line. A splitting of the jets into more jets is also possible. There is an
alternative way of looking at this process. The first generation quark and
antiquark are the ends of a colour string. As they fly away the string
stretches. After some time the string breaks. At the breaking point a quark
antiquark pair is formed, so that each of the pieces of the string is again a
string with a quark at one end and an antiquark at the other. The string pieces
break again and finally short strings appear, which go over into hadrons. It is
natural to expect that there is a time scale $\tau$ for the hadronization
process. Since, however, the system is highly relativistic, it is necessary to
specify in which frame this time should be measured. We choose the
centre-of-mass frame and assume that $\tau$ is the longitudinal proper time at
the creation of the hadron defined by

\begin{equation}
\tau = \sqrt{t^2 - z^2},
\end{equation}
where $t$ is the centre-or-mass time, when the hadron was created, and $z$ is
the corresponding coordinate measured along the jet direction (the transverse
dimensions are less important). An estimate of the velocity of the hadron is
$z/t$. These formulae have an interesting implication. Hadron production begins
at $t=\tau$ and first slow particles close to the interaction point are
produced ($|z|$ small). Only later and further from the interaction point (both
$t$ and $|z|$ large) do the fast hadrons appear. This production mechanism is
known as the inside-outside-cascade \cite{BJO}.

A very different multiple particle production process are the central
heavy-nucleus -- heavy-nucleus collisions, known also as central heavy ion
collisions. Here the usual picture is that of two spheres Lorentz-contracted
into coaxial discs --- we consider the centre-of-mass system --- penetrating
through each other. When the discs fly apart, many strings are simultaneously
stretched in a tube with a transverse radius of the order of the radii of the
colliding nuclei. For heavy nuclei and high energies the strings are so
numerous that they merge, e.g. into a quark gluon plasma. Then another poorly
understood process, known as freeze out, converts the plasma (or whatever is
the intermediate state) into hadrons.

There are many obvious questions to ask. What is the transverse radius of the
tubular (?) region, where the hadrons are created? One would expect about one
fermi or less for $e^+e^-$ annihilations and several fermi for heavy ion
collisions. What is the formation time $t$, which elapses between the moment of
collision and the moment, when the last hadron is produced directly? What is
the time $\Delta t$ between the direct production of the first hadron and of
the last? There are also many model dependent questions. In thermodynamical models
one asks about the temperature, in hydrodynamic models about the velocity of
the collective flow etc.

Let us review now the main results of the famous GGLP paper \cite{GGL}. Even if
one does not quite share G. Goldhaber's opinion quoted above, this is certainly
a very important paper and some familiarity with its content is necessary for
any discussion of the Bose-Einstein correlations in multiparticle production
processes.

\section{The GGLP contribution}

Consider two $\pi^+$-s with momenta ${\bf p}_1$ and ${\bf p}_2$ produced at
${\bf r}_1$ and ${\bf r}_2$. If the pions were distinguishable, the probability
amplitude to observe them them both at ${\bf r}$ would be a product of the
single particle contributions:

\begin{equation}
A_{Dis}({\bf r}) = \exp[i\phi_1 + i{\bf p}_1\cdot({\bf r_1} - {\bf r})]*
\exp[i\phi_2 + i{\bf p}_2\cdot({\bf r_2} - {\bf r})]
\end{equation}
In each of the square brackets, $\phi_i$ is the phase acquired by the particle
at birth and the other term is the phase accumulated while propagating from
${\bf r}_i$ to ${\bf r}$ with momentum ${\bf p}_i$. Since, however, the two
pions are identical bosons, it is mandatory to symmetrize the amplitude and a
more realistic formula is

\begin{eqnarray}
\label{AUND}
A_{Und}({\bf r}) = \frac{1}{\sqrt{2}}\exp[i(\phi_1+\phi_2) + i({\bf p_1 - p_2})
\cdot{\bf r}] \nonumber \\ \left[ \exp[i(\bf p_1\cdot r_1+p_2\cdot r_2)] +
\exp[i(\bf p_1\cdot r_2+p_2\cdot r_1)]\right].
\end{eqnarray}

Physically this means that we have added coherently the contribution
corresponding to the possibility that the two pions have exchanged their birth
points. Classically for different momenta and given birth points, if before the
exchange the two pions can reach the point ${\bf r}$, then in general after the
exchange they must miss it. One should keep in mind, however, that the distance
between the points ${\bf r}_1$ and ${\bf r}_2$ is of the order of a fermi,
while the distance between either of them and the point ${\bf r}$ is of the
order of a meter. Classically this would not help, but quantum-mechanically in
this situation the probability of reaching ${\bf r}$ by both pions is in the
two cases the same for all practical purposes. For comparison with experiment
the result should be averaged over all the possible pairs of points ${\bf r_1,
r_2}$. By averaging the amplitude (\ref{AUND}) nothing interesting is obtained.
Therefore, GGLP assumed that one should average the square of the absolute
value of the amplitude (\ref{AUND}). Physically this means that the
contributions from all the pairs of points ${\bf r_1, r_2}$ add incoherently.
Then the probability of finding a pair of pions with momenta ${\bf p_1, p_2}$
is

\begin{equation}
<|A_{Und}|^2> = 1 + <\cos[{\bf q}\cdot({\bf r_1-r_2})]>,
\end{equation}
where ${\bf q} = {\bf p_1 - p_2}$ and the Dirac brackets $<...>$ denote
averaging, with a suitable weight, over all the pairs of points ${\bf r_1,
r_2}$. Using as an example a Gaussian weight function

\begin{equation}
\rho({\bf r_1, r_2}) = (2\pi R^2)^{-3} \exp[-(r_1^2 + r_2^2)/(2R^2)],
\end{equation}
where $R$ is a constant with the dimension of length, GGLP found

\begin{equation}
\label{GAUSS}
<|A_{Und}|^2> = 1 + \exp[-{\bf q}^2 R^2].
\end{equation}

The parameter $R$ can be interpreted as the radius of the sphere, where the
pions are produced. Thus, finding $R$ from a fit to the experimental data
yields the size of the hadronization region. Note that formulae qualitatively
similar to (\ref{GAUSS}) hold for a broad class of weight functions. For ${\bf
q}^2 = 0$ the cosine being averaged equals one whatever are the points ${\bf
r_1}$ and ${\bf r_2}$. Thus the right hand side of (\ref{GAUSS}) for ${\bf q}^2
= 0$ must be equal two. For large values of ${\bf q}^2$, the cosine is a
rapidly oscillating function of the difference ${\bf r_1 - r_2}$. For smooth
weight functions, therefore, its average is very small and the right hand side
of (\ref{GAUSS}) equals approximately one. If the weight function contains only
one parameter with the dimension of length, let us denote it $R$, the width of
the region in ${\bf q}^2$ over which the right hand side of (\ref{GAUSS}) drops
from the value $2$ to, say, $1.5$ must be proportional to $R^{-2}$ for purely
dimensional reasons. Thus it is not difficult to improve over the Gaussian
Ansatz. There is a problem, however.

It is certainly not true that the probability of producing two identical pions
with momenta ${\bf p_1}$ and ${\bf p_2}$ depends only on the momentum
difference ${\bf q}$. GGLP could have tried to save the theory by introducing
some more complicated weight functions, which would depend also on the
momenta. They found, however, a much simpler and more brilliant solution. Note
that in the model the squared modulus of the amplitude for distinguishable
particles is a constant. Thus one may claim that the right hand side of
(\ref{GAUSS}) is not the two-particle probability distribution, but the ratio

\begin{equation}
C_2({\bf p_1,p_2}) = \frac{\rho({\bf p_1,p_2})}{\rho_{Dis}({\bf p_1,p_2})}.
\end{equation}
Here $\rho({\bf p_1,p_2})$ is the experimentally observed two-particle
distribution and $\rho_{Dis}({\bf p_1,p_2})$ is the distribution, which would
be observed, if the the identical pions were distinguishable, i.e. if the
Bose-Einstein symmetrization were switched off. Of course the numerator of this
expression cannot be measured, which has led to much discussion as described in
the following section. This is the famous normalization problem. GGLP chose
again a very simple solution. They substituted for $\rho_{Dis}$ the
distribution of $\pi^+,\pi^-$ pairs, where of course symmetrization does not
occur. The experimental results for the ratio $C_2$ could be fitted reasonably
well with formula (\ref{GAUSS}) and thus both the apparent attraction in
momentum space among pions of the same charge got explained and an estimate of
the radius $R$ was given.

Let us mention one more fruitful idea from the GGLP paper. The right-hand side
of formula (\ref{GAUSS}) depends on the Lorentz frame, where the momenta are
measured. Choosing the rest frame of the pair, where the energies of the two
pions are equal, we can rewrite the result in a covariant form:

\begin{equation}
C_2({\bf p_1,p_2}) = 1 + \exp[Q^2 R^2],
\end{equation}
where $Q^2$ is the square of the four-vector $p_1-p_2$.

\section{Normalization}

The problem of finding the best denominator for the function $C_2$ has
attracted much attention. Theorists often suggest to replace $\rho_{Dis}(\bf{
p_1,p_2})$ by the product of single particle distributions $\rho_1({\bf p_1})
\rho_1({\bf p_2})$. With this choice, the function $C_2$ becomes the familiar,
standard two-particle correlation function. Moreover, in many models terms
cancel between the numerator and the denominator making the formula simpler.
One may also notice that a two-particle density is normalized to $<n(n-1)>$,
while the single particle distribution is normalized to $<n>$. Thus a better
choice of the denominator might be to multiply the product of single particle
distributions by $<n(n-1)>/<n>^2$. The advantages and disadvantages of using
this factor have been recently discussed in \cite{MIV}. Since it does not
depend on $q^2$, usually it does not have much effect on the parameters of the
source found from fits. The identification of $C_2$ with the standard
correlation function, in spite of its advantages, is not very popular with
experimentalists. In order to explain the reason let us consider the following
simple example. Consider a high-energy reaction, where the two initial
particles go over into two well collimated jets. By momentum conservation the
two jets must be back to back in their centre-of-mass system. Let us assume
that the orientation of their common axis can point with equal probability in
any direction --- is isotropic. Then the single particle distribution of
momentum is also isotropic. On the other hand the opening angle between the
momenta of two particles is either small, when the two particles are extracted
from the same jet, or close to $\pi$ if the two particles are from different
jets. Thus the correlation function exhibits a very large peak for $\theta
\approx 0$. This peak has, of course, nothing to do with Bose-Einstein
correlations. Experimentalists would prefer a  definition of $C_2$, where the
forward peak reflects Bose-Einstein correlations and nothing else, because
this makes the interpretation much easier. Sometimes a compromise is chosen.
For instance in $e^+e^-$ annihilations, where the situation is similar to that
from our example, the $z$-axis is often chosen along the jet axis and not along
a direction fixed in the laboratory frame. With this choice the single particle
distribution becomes also strongly peaked for small angles $\theta$ and in the
correlation function the bumps due to the two-jet structure of the events are
largely eliminated.

The most popular choices, however, are improvements over the choice of GGLP.
For instance one uses "mixed" samples, where $\rho_{Dis}$ is the distribution
of pairs of $\pi^+$-ses, but with each $\pi^+$ taken from a different event.
Sometimes Monte Carlo generated samples are used, with Monte Carlo generators
which do not include Bose-Einstein correlations. This procedure is not very
safe, because such generators contain a number of free parameters, which are
fitted to the data, where the Bose-Einstein correlations are present. One also
uses
ratios of functions $C_2$ obtained from the data to functions $C_2$ obtained
according to the same prescription from Monte Carlo. The wide variety of
methods of calculating the denominator of the function $C_2$ is one of the
reasons, why the comparison of results from different experimental groups is
very difficult. This is, however only part of the story. Some groups correct
for final state interactions (mostly coulombic) and/or resonances, others do
not. Various cuts defining the data samples are used. Some groups assume that
every negative particle is a $\pi^-$, while other have particle identification.
Because of all that great care is necessary when interpreting the experimental
results and their stated errors. This difficulty has been known for a long time
cf. e.g. \cite{DEU}. For a more recent (pessimistic) review cf. \cite{HAY}.

\section{Beyond spherical symmetry}

The GGLP weight factor is a function of $r^2 = x^2 + y^2 + z^2$ and, therefore,
it is spherically symmetric. It is natural to replace $r^2$ by an arbitrary
quadratic form in $x,y,z$, provided the eigenvalues are positive so that the
weight function can be normalized to unity. Performing the averaging of the
cosine one obtains

\begin{equation}
C_2({\bf q,K}) = 1 + \lambda({\bf K})\exp[-\sum_{i,j = 1}^3 R_{ij}^2({\bf
K})q_iq_j].
\end{equation}

Here besides abandoning spherical symmetry two improvements have been
introduced. In agreement with experimental observation (cf e.g. \cite{DEU}) a
factor $0 < \lambda \leq 1$ has been included and a dependence of all the
coefficients on ${\bf K} = ({\bf p_1 + p_2})/2$ has been allowed. The coefficients denoted
$R^2_{ij}$ do not have to be all positive. Out of the nine coefficients
$R^2_{ij}$ three are eliminated by the symmetry condition $R^2_{ij} =
R^2_{ji}$. Moreover, choosing the $y$ axis so that $K_y = 0$ and assuming
reflection symmetry with respect to the $x,z$ plane we have $R^2_{yz} =
R^2_{yx} = 0$. Thus there are four independent coefficients left. The time
component $q_0$ can be easily obtained from the identity $K^\mu q_\mu =
(p_1^2 - p_2^2)/2 = 0$.

It is convenient to choose the $z$-axis in the longitudinal direction i.e. for
central heavy ion collisions along the beam axis and for $e^+e^-$ annihilations
along the jet axis. The $y$-axis is perpendicular to the $z$-axis and to the
vector ${\bf K}$. This fixes also the direction of the $x$-axis. With this
choice, the $x,y,z$ directions are often referred to as the {\it out} direction,
the {\it side} direction and the {\it longitudinal} direction respectively.
The
$R^2$
parameters are denoted $R_s^2, R_0^2, R_l^2$ and $R_{0l}^2$ \cite{BER,CSH}.
There were speculations that the study of $R_{out}/R_s$ could give clues as to
whether there is quark-gluon plasma and/or collective flow in the system
\cite{BER}, but the results have not been conclusive and a complete study of
all the parameters seems now to be the best strategy. Other choices of
parameters are also possible. For instance one can put

\begin{equation}
C_2({\bf q,K}) = 1 + \lambda\exp[-R_x^2 q_x^2 - R_y^2 q_y^2 -R_z^2 q_z^2 -T^2
q_0^2]
\end{equation}
\cite{MIS}, or

\begin{equation}
C_2({\bf q,K}) = 1 + \lambda\exp[-R^2_T q_T^2 -R_{\|}^2(q_L^2 - q_o^2) - (R_0^2
+ R_{\|}^2)\gamma^2(q_0^2-v q_z^2)],
\end{equation}
where

\begin{equation}
\gamma = \frac{1}{\sqrt{1 - v^2}},
\end{equation}
and $u = \gamma(1,{\bf 0_T},v)$. This parametrization proposed in \cite{YAK}
and improved in \cite{POD} is particularly popular and is often referred to as the
YKP parametrization.

\section{Time dependence}

In the GGLP picture the production of all the hadrons was instantaneous at some
time $t_0$. Formally this assumption is difficult to disprove. Choosing the
time $t_0$ after all the hadrons have been produced and interacted and before
the time when they were observed, one can calculate the distributions at the
observation time using the state at time $t_0$ as the initial condition.
Whether the hadrons existed before time $t_0$, is irrelevant for this
calculation. In particular, one may assume that all the hadrons were created at
time $t_0$. Guessing the initial condition at time $t_0$ is, however, very hard
in this approach. Therefore, it is more practical to choose a more realistic
conjecture about the origin of the hadrons, because then the initial conditions
are more natural and easier to guess.

In particular, several authors (cf. e.g. \cite{KOP,HAP,MAS}) assumed that the
production of hadrons is from sources, which fly away from each other and such
that in the rest frame of a source the production of hadrons is isotropic.
Consider the simplest case of just two sources. If their relative velocity is
large and the momenta of the hadrons in the rest frames of the corresponding
sources are moderate, then it is very unlikely that two identical pions from
different sources have momenta close to each other. On the other hand, all the
information about the structure of the source comes from pairs with small
momentum differences $|{\bf q}|$. For large values of $|{\bf q}|$ function
$C_2$ is flat and carries no information. Consequently, the observed function
$C_2$ contains only information about the single sources and no information
about the distance between them. One finds that the effective production region
is spherical in its rest frame, while the actual production region composed of
the two sources is elongated. This is an important piece of information. What
we observe is not the total size of the hadronization region, but the average
size of the so-called regions of homogeneity \cite{SIN}, i.e. regions, where
hadrons with similar momenta are produced. For this reason the effective
hadronization regions observed in experiment are approximately spherical,
while we expect that at high collision energy the actual production regions are
strongly elongated, because of their stringy origin.

There is one more interesting result connected with the problem of the time
span of the hadronization process \cite{BER}. Consider instantaneous
hadronization from a spherical shell of thickness $\delta R$. The length
$\delta R$ should be reflected in the momentum correlations along the direction
$x_{out}$, because the particles, which come from the inner part of the shell,
are out of phase with those, which come from the outer part of the shell. The
same effect can be obtained, however, if some particles are produced later
than others. From the experimental fact that $R_{out}$ is not particularly
large, one concludes that the hadronization process does not last very
long. This excludes models, where hadrons are produced from the quark-gluon
plasma in a first order phase transition with a large latent heat. This process
would be too slow to be made consistent with the observed moderate time
interval of the hadron production.

\section{More quantum physics}

The formulation of the GGLP model is quasiclassical -- one talks about a pion
with momentum ${\bf p}$ created at point ${\bf r}$, which is not quantum
mechanics. Much work has been done on formulations consistent with quantum
mechanics. When dealing with incoherent superpositions of states one should use
density matrices or density operators. The starting point may be the density
matrix in coordinate representation $\rho({\bf x},{\bf x'},t)$ or in momentum
representation $\rho({\bf p,p'},t)$. Often it is convenient to replace
the vectors ${\bf a},{\bf a}'$ by their linear combinations

\begin{equation}
{\bf a}_+ = \frac{1}{2}({\bf a} + {\bf a'});\;\;\;\;\;\;\;\;\;\;{\bf a}_- =
({\bf a} - {\bf a'})
\end{equation}
In particular, the GGLP results can be derived from the density matrix

\begin{equation}
\rho({\bf p},{\bf p'},t) = \int dr\langle {\bf p}|{\bf r}\rangle \rho({\bf r})
\langle {\bf r}|{\bf p'}\rangle.
\end{equation}
This incidentally shows that in spite of its pseudoclassical formulation the
GGLP model can be translated into respectable quantum mechanics.

The density matrices, however, do not combine explicitly the information about
the space distribution of sources and the momentum  distribution of the final
pions. Therefore, in order to derive the space distribution of sources from the
momentum distributions of the observed final particles other approaches have
been proposed.

One can use the Wigner function $W$ defined in term of the density matrix by
the formula

\begin{equation}
W({\bf p}_+,{\bf x}_+) = \int dp_- e^{i{\bf p}_-\cdot{\bf x}_+}\rho({\bf p},
{\bf p'},t).
\end{equation}
The properties of the Wigner functions are described in detail in the famous
review article \cite{HIL}. For a recent application to the description of
multiple production of identical particles cf. \cite{BIK}. In a well defined
sense \cite{HIL} Wigner's function is the best quantum mechanical analogue of
the classical phase space distribution. In this formulation Heisenberg's
uncertainty principle is easily implemented. The density in phase space should
not be too large. Quantitatively the condition is

\begin{equation}
|W({\bf p}_+,{\bf p}_-,t)|^2 \leq (\pi \hbar)^{-3n},
\end{equation}
where $3n$ is the dimension of the vectors ${\bf p}_\pm$, or equivalently $n$
is the number of particles described by the Wigner function $W$.

An alternative approach is to introduce the classical position and momentum
vectors ${\bf \xi}$, ${\bf \pi}$ besides the quantum mechanical position
and momentum vectors ${\bf x}$, ${\bf p}$. Each pair of vectors {\bf $\xi$},{\bf
$\pi$} defines a quantum mechanical wave packet e.g.

\begin{equation}
\langle{\bf x}|{\bf \xi}, {\bf \pi}\rangle =
\left(\frac{\sigma^2}{\pi}\right)^{\frac{3}{4}} \exp\left[\frac{1}{2}({\bf
x}
-
{\bf \xi})^2 + i{\bf \pi}\cdot{\bf x}\right],
\end{equation}
or equivalently in momentum representation

\begin{equation}
\langle{\bf p}|{\bf \xi}, {\bf \pi}\rangle =
\left(\frac{1}{\pi}\right)^{\frac{3}{4}} \exp\left[\frac{1}{2\sigma^2}({\bf p}
- {\bf \pi})^2 + i{\bf \xi}\cdot({\bf \pi}-{\bf p})\right].
\end{equation}
Coherent, or incoherent superpositions of such wave packets are legal quantum
mechanical states. On the other hand, the values of ${\bf x}$ and ${\bf p}$ are
constrained to be close to the values of ${\bf \xi}$ and ${\bf \pi}$. If the
distribution of the parameters ${\bf \xi}$ and ${\bf \pi}$ is calculated from
some classical model, the result is directly translated into a distribution of
the observables ${\bf x}$ and ${\bf p}$, which is consistent with quantum
mechanics. This approach was pioneered in ref. \cite{GKW}. For a recent
application and detailed discussion cf. \cite{ZIC}.

Still another approach is to introduce a source function $S$ related to
the density matrix by the formula

\begin{equation}
\rho(p,p') = \int d^4x \exp[ip_-x_+]S(x_+,p_+).
\end{equation}
In this formula $x,x',p,p'$ are fourvectors. The similarity of this formula to
the formula relating Wigner's function to the density matrix caused that the
source function is often referred to as a Wigner function, a kind of Wigner
function, a pseudo Wigner function etc. (cf. e.g. \cite{PRA}, \cite{CHH}). In
fact the relation between the source function and Wigner's function is not
unique and may be quite complicated. We will not discuss this problem here, but
let us note that the density matrices or Wigner's function were calculated at
a given time, while here the time dependence must be known. In simple cases the
time dependence of a Wigner function can be found and used (cf. e.g.
\cite{BGR}), but it is quite complicated and unlikely to be found by a simple
guess. What is more, the source function depends on two time arguments.

There is one case, however, when the source function has an easy
interpretation. If the pions are produced by classical currents $J$, then
\cite{SHU}, \cite{CHH}

\begin{equation}
S(x_+,p_+) = \int \frac{d^4x_-}{2(2\pi)^3}e^{-ip_+x_-}\langle
J^*(x)J(x')\rangle,
\end{equation}
where the averaging is over all the incoherent components of the currents
$J(x)$.

\section{Momentum-position correlations}

In the GGLP model and in many others the momentum spectrum of the produced
particles does not depend on the position of the source. When momentum-position
correlations are introduced in the model, unexpected results may be obtained.
Let us consider for example quasiclassical sources with the single particle
position-momentum distribution

\begin{equation}
\rho({\bf x},{\bf p}) = \sqrt{2\pi R^2}^3\delta({\bf p}-\lambda{\bf r})
\exp\left[-\frac{r^2}{2R^2}\right].
\end{equation}
Correlations of the type ${\bf p} \sim {\bf r}$ occur in classical versions of
various models, e.g. in string models and in models, where hadronization is
preceded by a rapid flow of hot matter. The average of the cosine from the
interference term in the present generalized GGLP model is

\begin{equation}
\langle\cos[{\bf q}\cdot({\bf r}_1 - {\bf r}_2)]\rangle = \lambda^6
\exp\left[-\frac{{\bf K}^2}{\lambda^2 R^2}\right]
\exp\left[-\frac{{\bf q}^2}{4\lambda^2R^2}\right] \cos\left[
\frac{{\bf q}^2}{\lambda}\right].
\end{equation}
In comparison with the GGLP result for the Gaussian weight function $\rho$, two
changes are striking. A dependence on the sum of momenta ${\bf K}$ appeared and
the dependence on the momentum difference ${\bf q}$ is no more Gaussian. For
non-Gaussian distributions the effective radius of the distribution of sources
in space is usually defined by

\begin{equation}
R^2_{eff} =
-\left(\frac{d}{d{\bf q}^2}\langle\cos[\ldots]\rangle\right)_{{\bf q}^2 = 0}.
\end{equation}
For the GGLP model with a Gaussian weight function this reproduces the usual
result $R_{eff} = R$. In the present case, however, we obtain

\begin{equation}
R_{eff} = \frac{1}{2\lambda R}.
\end{equation}
Due to the position-momentum correlations, $R_{eff}$ becomes inversely
proportional to $R$!

A more realistic model with similar correlations can be built as follows
\cite{CSZ},\cite{BZ1}. For the source function we choose

\begin{equation}
S(x,p) = \delta(t^2-z^2-\tau_0^2) \exp\left[ \frac{(p-\lambda x)_+(p-\lambda
 x)_-}{2\delta_\parallel^2}  - \frac{({\bf p}-\lambda
{\bf x})_T^2}{2\delta_T^2}
- \frac{{\bf x}_T^2}{2R^2}\right].
\end{equation}
This formula has a simple physical interpretation. The subscript $T$ denotes
the vector component transverse with respect to the beam axis $z$. The last
term in the exponent implies that the particles are created not too far from
the beam axis, typical distances being of the order of the parameter $R$. The
first two terms in the exponent impose the condition ${\bf p} \approx \lambda
{\bf x}$. The subscripts $\pm$ refer to the fourvectror components $a_0 \pm
a_z$. The dispersion of the transverse components is of the order of
$\delta_T^2$ and that of the longitudinal component of the order of
$\delta_\parallel^2$. The Dirac delta implies that for each drop of the hot
matter hadronization takes place after the same longitudinal invariant time
$\tau_0$. It is useful that for this source function the corresponding density
matrix in the momentum representation can be calculated in closed form. One
finds in particular

\begin{equation}
|\rho({\bf K},{\bf q})| \sim \left(\delta_T^2 +
\frac{R^2M_T^2}{\tau_0^2}\right)
\exp\left[-\frac{{\bf K}^2_T + R^2\delta_T^2 {\bf q}_T^2}{\delta_T^2 +
R^2M_T^2\tau_0^{-2}}\right] \exp\left[\frac{2M_T^2}{\delta_\parallel^2}\right]
|K_0(\alpha)|^2,
\end{equation}
where $K_0$ is the modified Bessel function,

\begin{equation}
\alpha^2 = \frac{M_T^4}{\delta_\parallel^4} -
\frac{\tau_0^2({\bf K}_T\cdot{\bf q}_T)^2}{M_T^2} - \frac{\tau_0^2m_{1T}^2
m_{2T}^2}{4M_T^2}\sinh^2(y_1-y_2) + \frac{2 i \tau_0 M_T
{\bf K}_T{\bf q}_T}{\delta_\parallel^2},
\end{equation}
and $m_{iT}^2 = m_{\pi}^2 + {\bf p}_{iT}^2$ and $M_T^2 = K^2 + {\bf K}_T^2$ are
transverse masses of the two particles and of the pair. The single particle
distribution can be calculated from the formula $\rho_1({\bf p}) =
\rho({\bf p},0)$.

This model was found to reproduce reasonably well the data for single particle
distributions and for Bose-Einstein correlations in $e^+e^-$ annihilations at
LEP energies \cite{BKP}.

\section{Final state interactions and (partial) coherence}

In the GGLP model and in many later models, after hadronization the pions
propagate as free particles. In fact we know, that the majority of pions is
generated in resonance decays. Moreover, there are the nonresonant strong and
the electromagnetic interactions among the pions. All that has been studied for
years, but many problems are still controversial.

Strong nonresonant interactions are usually neglected, but it has been pointed
out \cite{PRA} that absorption of the produced pions can lead to a decrease of
the effective radius $R_{eff}$ with increasing momentum of the pair
$|{\bf K}|$. The effect of resonances seems to be much more important. The long
lived resonances, mostly $\eta$ and $\eta'$, usually decay far from the centre
of the interaction region. Therefore they simulate a large hadronization region
and produce a narrow peak in the plot of $C_2$ versus ${\bf q}^2$. This peak is
narrower than the experimental resolution and consequently its main effect is
to reduce the parameter $\lambda$. The resonances with life times neither very
long, nor very short, like the $\omega$ resonance, produce for small
${\bf q}^2$ a steep rise of $C_2$ with decreasing ${\bf q}^2$. Short lived
resonances, like the $\rho$ meson, increase only slightly the measured radius
of the interaction region. It has been suggested \cite{BOW}, \cite{HAY} that in
$e^+e^-$ annihilation it may be difficult to explain why after correcting for
the resonance effects the Bose-Einstein correlations remain as strong as
observed in experiment.

Coulomb interactions can be easily included by replacing in the description of
the propagation of the pions the plane waves by Coulomb wave functions, which
leads to the introduction of the so called Gamow factor \cite{GKW}. This is now
known to grossly overestimate the effect (for a review cf. \cite{BAY}). The
physical reason is that the introduction of the Coulomb wave functions
describes the evolution of an isolated pair of pions, while in reality there are
many other pions around, which partly screen the Coulomb interaction within
the pair. A direct experimental argument is that the reasoning leading to the
Gamow factor, when applied to $\pi^+\pi^-$ pairs, gives a strong attraction,
which is not seen in the data. A simple way of correcting for the screening
effect is given in ref \cite{BBM} (see also \cite{BAY}). A screening radius
$r_0$ is introduced. The interaction potential of the pair is continuous at
$r=r_0$, Coulomb for $r>r_0$ and constant for $r<r_0$. The parameter $r_0$ is
chosen so as to reproduce correctly the experimental data for $\pi^+\pi^-$
pairs. Coulomb corrections calculated in this way are rather small.

Another assumption of GGLP, which has been put into doubt, was that the
production process is completely incoherent. Complete coherence would kill the
effect, but some degree of coherence seems difficult to avoid in realistic
models. It is easy to write down general formulae including partial coherence
(cf. e.g. \cite{GKW}), but it is not clear how to use them fruitfully. At a
time it was suggested that the parameter $\lambda$ measures the degree of
coherence ($\lambda = 1$ no coherence, $\lambda = 0$ complete coherence), but
now it is clear that this parameter is strongly affected by the dynamics of the
production process, in particular by resonance production, and by experimental
conditions (e.g. particle misidentification).

\section{Two recent models}

In order to illustrate how Bose-Einstein correlations are nowadays analysed, we
present two recent models. The first (cf. \cite{WIH} and references quoted
there) is based on analogies with hydrodynamics and thermodynamics. It is
being used to describe central heavy ion collisions. The starting point is the
(single particle) source function

\begin{eqnarray}
S(x,p) \sim m_T\cosh(y-\eta) \exp\left[\frac{m_T\cosh y\cosh\eta_T(r_T)-
p_T\frac{x}{r}\sinh\eta_T(r_T)}{T}\right] \nonumber \\
\exp\left[-\frac{r_T^2}{2R^2} -
\frac{\eta^2}{2(\Delta\eta)^2} -
\frac{(\tau-\tau_0)^2}{2(\Delta\tau)^2}\right].
\end{eqnarray}
In this formula $\sim$ means that a normalization constant has been omitted,
$y$ is the rapidity along the $z$ axis i.e. along the beam direction. The
pseudorapidity

\begin{equation}
\eta = \frac{1}{2}\ln\frac{t+z}{t-z}.
\end{equation}
The distance from the $z$ axis $r_T = \sqrt{x^2+y^2}$. The rapidity of the
transverse flow has been assumed in the form

\begin{equation}
\eta_T(r_T) = \eta_f\frac{r_T}{R},
\end{equation}
where $\eta_f$ is a constant. The model contains six free parameters: $R, T,
\eta_f, \Delta\eta, \tau_0$ and $\Delta\tau$. These parameters have been
estimated by comparison with the data from the NA49 experiment for collisions
of lead nuclei at an energy of 158 GeV per nucleon. Some of the results have
interesting physical implications. The parameter $R$, interpreted as the
transverse radius of the hadronization region, is about 7 fm. From the known
size of the lead nucleus one could have expected a number about twice smaller.
This means that there is substantial transverse spreading of the interaction
region before hadronization takes place. The parameter $T$ interpreted as the
local temperature is about 130 MeV. This is less than the temperatures found in
models used to calculate the chemical composition of the produced hadrons,
which could mean that the spreading is accompanied by cooling. The parameter
$\eta_f$ is about 0.35. This is a very reasonable value. The velocity of sound
in a plasma is about 1/3 (in units of the velocity of light in vacuum). The
parameters $\tau_0$ and $\Delta\tau$ are about 9fm and 1.5fm respectively. It
means that the time span of hadronization is short compared to the time between
the original interaction and the onset of hadronization. This would exclude
models, where hadronization is a first order phase transition with a large
latent heat, because in such models the time span of hadronization is large.
One should keep in mind, however, the authors' warning that the determination of
the parameter $\Delta\tau$ from the data is poor. This example shows that given
a model one can extract from the data much interesting information. Little is
known, however,  about the model dependence of these results.

A very different picture of the Bose-Einstein correlations \cite{ANH,ANR,AN1,AND}
has been inspired by a string model of the Lund type. This model is
taylored for $e^+e^-$ annihilations. An annihilation is depicted as the
formation of a string with a quark at one end and an antiquark at the other.
This string stretches with the speed of light. Then somewhere along the string
a quark-antiquark pair is produced and the string breaks. The pieces stretch
and break again. Finally sufficiently short bits of strings hadronize. In the
$z,t$ plane the trajectories of the ends of all these strings form a closed
contour. Let us denote the area enclosed by this contour by $A$. The key
assumption is that the probability amplitude for a given final state is

\begin{equation}
M \sim \exp[i\xi A].
\end{equation}
We have not written explicitly the factor related to the transverse momenta of
the produced particles. Denoting by $b/2$ the imaginary part of $\xi$ one finds

\begin{equation}
|M|^2 \sim e^{-bA}.
\end{equation}
This result is well known from the Lund model. For the description of the
Bose-Einstein correlations, however, it is the real part of $\xi$ which is the
important one. It is expected to be of the order of the string tension i.e. of
the order of 1 GeV/fm. The point is that the production amplitude has to be
symmetrized with respect to exchanges of identical particles. Such an exchange,
however, changes the area $A$ and because of the nonzero real part of $\xi$
the phase of the amplitude. Numerical calculations show that this model gives a
reasonable description of the Bose-Einsten correlations in $e^+e^-$
annihilations. This is very interesting, because this model, contrary to all
the previous ones, does not contain incoherent components, random phases etc.
It is curious what happens, when in a process more than one string is
initially produced. In $e^+e^-$ annihilations there are events, where two $W$
bosons are produced, which implies two strings. It has been suggested
\cite{AND} that in this case only pions from a single string should exhibit
Bose-Einstein correlations. To be sure, nobody doubts that pions are bosons and
that consequently their production amplitude should be suitably symmetrized,
but the attraction in momentum space, known since the GGLP paper as
Bose-Einstein correlations, requires in addition certain phase relationships,
which may not be realized for pions originating from different strings.
In central heavy ion collisions many strings are produced. By extension of the
previous argument one could expect very weak Bose-Einstein correlations, which
experimentally is not the case. This suggests that the state from which
hadronization occurs in heavy ion collisions is not a bunch of strings. The
strings must somehow merge and form a very different object, perhaps a volume
of quark-gluon plasma.

\section{Conclusion}

Bose-Einstein correlations attract much interest. Many hundreds of papers have
been published on this subject. In principle they offer the only access to some
important information about the hadronization process. Typical questions are:
what is the size and shape of the hadronization region, how is hadronization
distributed in time, what is the coherence of the sources, what is the
hadronization mechanism etc. In practice it is difficult to find model
independent, definitive answers to these questions. The common approach is to
study models, which either demonstrates the viability of certain scenarios,
like the sting model discussed above, or give tentative answers to the
questions concerning hadronization.

\end{document}